\documentclass{article}

\usepackage{arxiv}
\usepackage[utf8]{inputenc} 
\usepackage[T1]{fontenc}    
\usepackage{hyperref}       
\usepackage{url}            
\usepackage{booktabs}       
\usepackage{amsfonts}       
\usepackage{nicefrac}       
\usepackage{microtype}      
\usepackage{lipsum}
\usepackage{amsmath,amssymb}

\usepackage{float}
\usepackage{multirow}
\usepackage{algorithm,algpseudocode}
\usepackage[algo2e]{algorithm2e}
\SetKwRepeat{Do}{do}{while}
\DeclareMathOperator*{\argmin}{arg\,min}

\usepackage{graphicx}
\usepackage{array}
\usepackage{tabu}

\title{Photo-acoustic tomographic  image reconstruction from reduced data
	using physically  inspired regularization}

\author{
	Nadaparambil Aravindakshan Rejesh \\
	Imaging Systems Lab\\  Department of Electrical Engineering\\
	Indian Institute of Science (IISc)\\ Bangalore 560012, India\\
	\texttt{rejeshn@iisc.ac.in} \\
	\And
	Sandeep Kumar Kalva \\
	School of Chemical and Biomedical Engineering\\ 
	Nanyang Technological University\\ 62 Nanyang Drive\\ 
	637459 Singapore \\
	\texttt{sandeepk002@e.ntu.edu.sg} \\
	  \And
	Manojit Pramanik \\
	School of Chemical and Biomedical Engineering\\ 
	Nanyang Technological University\\ 62 Nanyang Drive\\ 
	637459 Singapore \\
	\texttt{manojit@ntu.edu.sg} \\
	 \And
  Muthuvel Arigovindan
  \thanks{Corresponding author}    \\
  Imaging Systems Lab\\ Department of Electrical Engineering\\
  Indian Institute of Science (IISc)\\ Bangalore 560012, India\\
    \texttt{mvel@iisc.ac.in} \\
}
\begin{document}
\maketitle

\begin{abstract}
We propose a model-based image reconstruction method for photoacoustic tomography (PAT) involving a novel form of regularization and demonstrate its ability to recover good quality
images from significantly reduced size datasets. The regularization is constructed to suit the physical structure of typical PAT images. We construct it by combining second-order derivatives and
intensity into a non-convex form to exploit a structural property of PAT images that we observe: in PAT images, high intensities and high second-order derivatives are jointly sparse. The specific form
of regularization constructed here is a modification of the form proposed for fluorescence image restoration. This regularization is combined with a data fidelity cost, and the required image
is obtained as the minimizer of this cost. As this regularization is non-convex, the efficiency of the minimization method is crucial in obtaining artifact-free reconstructions. We develop a custom
minimization method for efficiently handling this non-convex minimization problem. Further, as non-convex minimization requires a large number of iterations and the PAT forward model in the
data-fidelity term has to be applied in the iterations, we propose a computational structure for efficient implementation of the forward model with reduced memory requirements. We evaluate the proposed method on both simulated and real measured data sets and compare them with a recent reconstruction method that is based on a well-known fast iterative shrinkage threshold algorithm
(FISTA).
\end{abstract}


\section{Introduction}

Photoacoustic tomography (PAT) \cite{wang2016practical,pramanik2008design,beard2011biomedical,
	zhou2016tutorial,upputuri2016recent,li2017single, Han:17}  provides high resolution and 
high contrast images of deep tissues by imaging the spatial distribution  of  certain  
substances   that can absorb near infra-red optical energy.  
Upon shining with a laser pulse,  the substance under investigation
absorbs the optical energy and undergoes thermoelastic expansion; thus,   the spatial distribution of the concentration of the substance 
gets translated into the distribution of pressure-rise.   This initial pressure rise 
travels outwards as ultrasound waves which are collected by ultrasound 
transducers placed at the boundary. 
From the ultrasound signal measured by the transducers
as a function of time, a PAT reconstruction method recovers an estimate
of the initial pressure-rise by solving the associated inverse problem 
\cite{rosenthal2013acoustic}.  Achieving
accuracy in image reconstruction  is a fundamental challenge in  
PAT;  the reconstructed images   suffer from artifacts that depend on the 
measurement geometry mainly because  the reconstruction problem is
complex.

There is a class of methods known as analytical inversion methods, which obtain reconstruction (recovery of initial pressure rise) by some transformation on the measured data.  Among such methods, filtered back projection (FBP) \cite{finch2004determining, kunyansky2007explicit, xu2002time, xu2005universal} and delay and sum method \cite{kalva2016experimental, pramanik2014improving} is fast and memory-efficient but require a large number of measurements for good quality reconstruction.  This leads to increased scan time or expensive instrumentation setups. Time reversal methods \cite{xu2004time}  are the least demanding class of methods in the category of analytical methods and can be used for any detection geometry \cite{burgholzer2007exact}, acoustic absorption, and dispersion \cite{treeby2010photoacoustic}, and can accommodate the heterogeneities present in the medium \cite{hristova2008reconstruction}. 
However, in many applications of PAT, due to the geometrical limitations \cite{xu2004reconstructions} or by choice to accelerate data acquisition \cite{arridge2016accelerated}, we may have to put restrictions on the spatial and/or temporal sampling of the photoacoustic (PA) signal. In such situations, these methods suffer from blurring and curved-line artifacts.

Model-based reconstruction methods outperform the direct analytic reconstruction methods in the case of limitation in the size of measured data and yield better quality in reconstruction
\cite{buehler2011model, paltauf2002iterative, schwab2018galerkin, jp2019}. These methods express the measured pressure samples as a linear transformation on the initial pressure-rise.  This transformation is implemented in the form of multiplication of the vector containing image pixels with a very large matrix that represents the PAT forward model (model matrix).   This transformation also goes into the data fidelity term, which is minimized along with a regularization functional to achieve reconstruction.  The regularization represents a prior belief on the spatial characteristics of the image to be recovered  \cite{huang2013full, arridge2016adjoint, boink2018framework,
	saratoon2013gradient}.  The minimization is achieved using iterations involving the repeated application of the above-mentioned transformation and its adjoint leading to high computational burden and memory-burden.  To reduce memory requirements and computational complexity,  several methods have been reported. For example, the memory overhead can be reduced by decreasing the number of measurements  
\cite{han2015sparsity} or by calculating the matrix-vector products on-the-fly without explicitly storing the model matrix \cite{wang2013accelerating}. However, since the same operations need to be performed multiple times, the computational time becomes longer.  A fast implementation can be achieved by simplifying the forward model, assuming that the photoacoustic sources lie in a plane \cite{rosenthal2010fast}.  It can also be made faster by decoupling the inversion into smaller subproblems using a discrete wavelet packet decomposition \cite{rosenthal2012efficient}. However, the method uses interpolation, which leads to modeling errors. 
A more accurate model matrix calculation based
on the direct discretization of Poisson-type integral showed better 
reconstruction accuracy in high-noise and low-SNR imaging conditions \cite{liu2016curve}.

We develop a novel model-based reconstruction method that yields significantly improved reconstruction from datasets of reduced
size.  Our contributions are threefold. First,  we note that the
existing model-based methods in PAT use generic regularization forms developed for general images.   There has been no reported application of a regularization that can cater to the structural properties of the photoacoustic images. We observe that PAT images have the following characteristics, which are also observed in fluorescence images:  in these images, high intensities and high second-order derivatives are jointly sparse. This property was exploited for fluorescence image reconstruction by constructing
a regularization that combines the intensity and the second-order derivatives  \cite{arigovindan2013high}. Here we modify this form such that it is more suitable for the current reconstruction problem.   
Second,  we construct a novel preconditioned gradient method for efficiently minimizing the cost function that combines this regularization and the data-fidelity term. Finally, we develop an efficient method for on-the-fly computation of matrix-vector products derived using a model that uses an exact time propagator to calculate the acoustic field \cite{cox2007k,cox2005fast}. By taking advantage of the filtering structure present in this formulation, forward and adjoint matrix operations are  
implemented using filters in a memory-efficient way. 
We evaluate the proposed method and compare it with a recent reconstruction method based on a well-known fast iterative shrinkage threshold algorithm (FISTA).
For this purpose, we use simulated data corresponding to the reduced number of transducers and real measured data.

\subsection{Forward model in photoacoustic tomography}
The forward problem in PAT accounts for the calculation of the pressure fields in 
space and time $p(\textbf{r},t)$ from a known photoacoustic source $H(\textbf{r},t)$, 
which represents the light energy deposited in the medium per unit volume per unit 
time. The induced pressure waves $p(\textbf{r},t)$  under the condition of thermal 
and stress confinements obey the following differential equation for an acoustically 
homogeneous medium \cite{rosenthal2013acoustic,zhou2016tutorial,xu2006photoacoustic}
\begin{equation}
	\label{eq:waveeqn}
	\Big( \frac{\partial^{2} }{\partial t^{2}} - c^{2}_{0} \; \nabla^{2}  \Big) \; p(\textbf{r},t) =  \Gamma \; \frac {\partial }{\partial t} H(\textbf{r},t),
\end{equation}
where $\Gamma$ is a dimensionless parameter called the Gr$\ddot{u}$neisen coefficient, which describes the conversion efficiency of heat to pressure and $c_{0}$ is the speed of sound in the medium. By recognizing that the temporal duration of the laser pulse is shorter than the temporal resolution of the ultrasound detectors in most of the photoacoustic imaging applications, the PA source $H(\textbf{r},t)$  may be approximated by  $H(\textbf{r})\delta(t)$, where $H(\textbf{r})$  is the density of deposited energy. Then the solution to the differential equation 
\eqref{eq:waveeqn}  can be written as \cite{rosenthal2013acoustic}

\begin{equation}
	\label{eq:contfmodel}
	p(\textbf{r},t) = \cfrac{\Gamma}{4 \pi c_{0} } \;\; \frac{\partial }{\partial t}\; \int\limits_{ |\textbf{r} - \textbf{r}^\prime|= c_{0} t} \; 
	\dfrac{H(\textbf{r}^\prime)}{|\textbf{r} - \textbf{r}^\prime|}\; d\textbf{r}^\prime,
\end{equation}
where the initial pressure field ${p}_0(\textbf{r})= p(\textbf{r},t=0)$ can be written as
\begin{equation}
	{p}_0(\textbf{r})= \Gamma \;H(\textbf{r}).
\end{equation}
The pressure distribution, $p(\textbf{r},t)$, can also be  expressed as
\cite{cox2005fast,cox2007k} 
\begin{equation}
	\label{eq:coxexp}
	p(\textbf{r},t) = \mathcal{F}^{-1} \Big\{ \hat{P}_{o}(\textbf{k}) \cos (c_{0}\|\textbf{k}\|t)\Big\}, 
\end{equation}
where  $\mathcal{F}^{-1}$ denotes the Fourier inversion,  and   $\hat{P}_{o}(\textbf{k})$ is the Fourier transform 
of ${p}_0(\textbf{r})$ with 
$\mathbf{k}$ denoting the 2D Fourier frequency.

\subsection{Discrete forward model}
The discrete representation of the forward problem is the basis of model-based reconstruction algorithms in PAT. Since the measured pressure field is linearly related to the photoacoustic source, the discretization of the forward problem may be written in a matrix form as \cite{paltauf2002iterative}
\begin{equation}
	\textbf{p}_m = \textbf{H} \textbf{p}_0 \;,
	\end {equation} 
	where $\textbf{p}_m$ is a  $LM \times 1$  vector representing the  
	discrete pressure measurements  from $L$ transducers each taking $M$ time-samples and $\textbf{H}$ is the model matrix of size $LM \times N$ with $N$ being the total number of pixels.   The initial pressure  distribution $p_0(\textbf{r})$ is represented  spatially using a 2D imaging grid having $N_{x}$ and $N_{y}$ grid points along $x$ and $y$ directions respectively  and  is denoted by a $N \times 1$ vector $\textbf{p}_0$ where $N =N_{x} \times N_{y}$. The model matrix can be calculated by discretizing the integral relation in \eqref{eq:contfmodel} in the ideal case of homogeneous lossless medium and point detectors. Several methods to improve the accuracy of the model by incorporating transducer responses, heterogeneity  in the medium and interpolation techniques for accurate discretization has been reported \cite{huang2013full, wang2011imaging, rosenthal2010fast}.  Including the measurement noise, the modified imaging model can be written as
	\begin{equation}
		\textbf{p}_m = \textbf{H} \textbf{p}_0 + {\pmb \eta},
		\end {equation} 
		where ${\pmb \eta}$ represents measurement noise, which is Gaussian. 
		
		\subsection{Model based image reconstruction}
		
		A reconstruction method in PAT aims to recover the initial pressure distribution $\textbf{p}_0$ from the noisy transducer measurement data $\textbf{p}_m$. In the limited data case, the PA image  reconstruction problem is ill-posed, and hence constraints
		are imposed on the required solution in the form of regularization. 
		In this case, the image  reconstruction problem can be treated as an optimization problem where the solution is  obtained by minimizing a cost function. The reconstruction problem can be written as
		\begin{equation}
			\hat{\textbf{p}}_0 = \argmin _{\textbf{p}_0} \; J(\textbf{p}_0)
		\end{equation} where $J(\textbf{p}_0)$ is the cost function and is given by 
		\begin{equation}
			J(\textbf{p}_0) = || \textbf{p}_m - \textbf{H} \textbf{p}_0 ||_{2}^{2} + \lambda \; {R}(\textbf{p}_0).
		\end{equation}
		Here ${R}(\textbf{p}_0)$ is the regularization functional and $\|\cdot\|_{2}$ represents the $l_{2}$ norm.  
		The regularization parameter $\lambda$ controls the amount of regularity in the solution and fidelity to the measured data.
		The regularization functional typically should be able to smooth the noise and hence contains derivative terms in their formulation.  A quadratic and differentiable regularization functional called Tikhonov regularization has been used in limited data cases  
		\cite{rosenthal2013acoustic, wang2012investigation} and is given by
		\begin{equation}
			{R}(\textbf{p}_0) =\sum_{i}\| \textbf{D}_{o, i}\textbf{p}_{0}\|^{2}_{2} = \sum_{r=1}^{N} 
			\sum_{i}\big((\textbf{D}_{o,i}\textbf{p}_0)_r\big)^{2}
		\end{equation} 
		where $(\cdot)_r$ denotes the $r$th component of its vector argument, and $\textbf{D}_{o,i}$ represents the matrix of $i^{th}$ derivative filter of order $o$.  For example, $\textbf{D}_{1,i} , i=1,2$, are the matrix equivalents of filtering by discrete filters that implement the operators $\frac{\partial}{\partial x}$ and  $\frac{\partial }{\partial y}$.   Further, $\textbf{D}_{2,i} , i=1,2,3$, are the matrix equivalents of filtering by discrete filters that implement the operators $\frac{\partial^{2} }{\partial x^{2}}, \frac{\partial^{2} }
		{\partial y^{2}}$  and $\sqrt{2}\frac{\partial }{\partial x \partial y}$.
		The resulting minimization of the convex quadratic cost function yields a closed form solution given by
		\begin{equation}
			\textbf{p}_{0} = \big[ \textbf{H}^{T}\textbf{H} + \lambda \;\sum_{i}\textbf{D}_{o,i}^{T}\textbf{D}_{o,i}\big] ^{-1} \textbf{H}^{T} \textbf{p}_m.
			\label{Tikh}
		\end{equation} 
		As it precludes any large derivative values,  Tikhonov regularization tends to smooth edges in the reconstructed image.
		
		It should be emphasized that, as of now,  only the data-fidelity term requires building a large matrix ${\bf H}$, and the regularization
		does not require building the matrices.  The matrices,  $\textbf{D}_{o, i}$s,  in the regularization  correspond to 
		discrete filters implementing the derivatives 
		$\frac{\partial}{\partial x}$,  $\frac{\partial } {\partial y}$,   $\frac{\partial^{2} }{\partial x^{2}}$,  $ \frac{\partial^{2} }
		{\partial y^{2}}$  and $\sqrt{2}\frac{\partial }{\partial x \partial y}$.  These filters can be directly applied to the images without building
		the matrices, which is more efficient.  We use the matrix formulation for derivatives also for achieving notational convenience
		in describing the minimization method. In the latter part of our development, we will eliminate the need for building
		the matrix ${\bf H}$ as well.

		A widely used non-quadratic regularization  is the  Total Variation (TV)  
		\cite{tao2009alternating, arridge2016accelerated, wang2008new, huang2013full} 
		and it  is superior to the quadratic functional in its ability to preserve edges and it is
		robust to noise. The discrete total variation is given by
		\begin{equation}
			{R}_{TV}(\textbf{p}_{0})  = \sum_{r=1}^{N}  \sqrt{\sum_{i}\big(\textbf{D}_{o,i}
				\textbf{p}_{0}\big)_{r}^{2}}
		\end{equation} 
		Often a differentiable   approximation of the total 
		variation is used which can be written as
		\begin{equation}
			{R}_{TV}(\textbf{p}_{0})  = \sum_{r=1}^{N}   \sqrt{\epsilon 
				+ \sum_{i}\big(\textbf{D}_{o,i}\textbf{p}_{0}\big)_{r}^{2}}
		\end{equation} 
		where $\epsilon$ is a small positive number.

		\section{The proposed reconstruction method}
		
		\subsection{Proposed Regularization functional}

		The photoacoustic images have high contrast due to the differential absorption of light in the near-infrared region by chromophores such as hemoglobin. Due to this,  high values of initial pressure, $p_0(\mathbf{r})$, are sparsely distributed.  Further,  regions having high derivative values are also sparsely distributed.  This pattern is also observed for fluorescence images in the work presented in \cite{arigovindan2013high},   where the regularization was constructed by adding an intensity term to second-order derivatives.  The combined point-wise cost went into a logarithmic function and summed over all pixels.  Here we replace the log by a fractional power and write the proposed regularization as
		\begin{equation}
			\label{eq:rh1}
			{R}_{h,1}(\textbf{p}_{0}, q)  = \sum_{r=1}^{N}   \left(\epsilon + \alpha \big(\textbf{p}_{0}\big)_{r}^{2} + (1-\alpha)\sum_{i}\big(\textbf{D}_{o,i}\textbf{p}_{0}\big)_{r}^{2}\right)^q,\;\; s.t. \;\;  0 < \alpha < 1,
		\end{equation} 
		where the weight $\alpha \in (0,1)$ controls the relative penalization.   The advantage of this modification  is that it allows an optimization strategy that can efficiently handle non-convex cost function, which will be demonstrated later.  Here, we choose $q < 0.5$ meaning that the resulting cost functional  is non-convex.  
		We also consider a variant of the above form, which is given below:
		\begin{equation}
			\label{eq:rh2}
			{R}_{h,2}(\textbf{p}_{0}, q) = \alpha \sum_{r=1}^{N}   \left(\epsilon +  \big(\textbf{p}_{0}\big)_{r}^{2} \right)^q  
			+ (1-\alpha)  \sum_{r=1}^{N}   \left(\epsilon + \sum_{i}\big(\textbf{D}_{o,i}\textbf{p}_{0}\big)_{r}^{2}\right)^q.
		\end{equation} 
		For notational convenience in describing the minimization algorithm, we call $q$ the sparsity index.
		
		\subsection{The complete cost functional}

		The initial pressure distribution in PAT is proportional to the fluence distribution and absorption coefficient in the tissue and hence the recovered PAT images should contain only non-negative values in it. Therefore, a non-negativity constraint is imposed on the solution of the optimization problem. The modified optimization problem to be solved is given by
		\begin{equation}
			\label{eq:fullprob}
			\hat{\textbf{p}}_{0} = \argmin _{\textbf{p}_{0}\geq 0} \; J(\textbf{p}_{0}, q)
		\end{equation} 
		where
		\begin{equation}
			J(\textbf{p}_{0}, q) = || \textbf{p}_m - \textbf{H} \textbf{p}_{0} ||_{2}^{2}  +
			\lambda R_{h,l}(\mathbf{p}_0, q)\;\;\; l=1\; \mbox{or} \; 2.
		\end{equation} 
		This constrained optimization problem can be efficiently handled by splitting-based optimization methods.  However, such methods have slow convergence and yield reconstructions with artifacts when the regularization is non-convex. Here,  we intend to develop a method based on traditional gradient-based iterative schemes,  and for this purpose,  we introduce an  approximate unconstrained
		formulation.  Specifically,  we modify cost by adding quadratic penalty term to enforce positivity as given below:
		\begin{equation}
			\label{eq:ucprob}
			J(\textbf{p}_{0}, q) = || \textbf{p}_m - \textbf{H} \textbf{p}_{0} ||_{2}^{2}  +
			\lambda R_{h,i}(\mathbf{p}_0, q) +\lambda_p \|\mathcal{P}^{-}(\textbf{p}_{0}\big)\|_2^2
		\end{equation} 
		where 
		\begin{equation}
			( \mathcal{P}^{-}(\textbf{p}_{0}))_{r} =
			\begin{cases}
				0 \;\;\;  &\text{if} \;\;   ( \textbf{p}_{0})_{r} \geq 0  \\
				(\textbf{p}_{0})_{r}  \;\;\;  &\text{if} \;\;   ( \textbf{p}_{0})_{r} < 0\; .
			\end{cases}
		\end{equation}
		Note that the problems in the equations (\ref{eq:fullprob}) and (\ref{eq:ucprob}) are equivalent only when $\lambda_p$ is arbitrarily large.  However,  setting  $\lambda_p$ to an arbitrarily large value will lead to numerical  instability.  Fortunately, through a
		series of reconstruction trials, we found that setting $\lambda_p=10\lambda$ was sufficient to avoid significant
		negative values.

		\subsection{Proposed Algorithm}
		
		Here, we first develop the minimization algorithm for reconstruction with  $R_{h,1}$ and then describe the modifications necessary to use $R_{h,2}$. We adapt the preconditioned gradient search for minimizing the cost function.  To this end, we first need to write the expression of the gradient.  For notational convenience,
		we use $\mathbf{x}$ in the place of  $\mathbf{p}_0$.  The gradient expression is given by 
		\begin{equation}
			\label{eq:graddef}
			\nabla J({\bf x}, q) =  \mathbf{A}^{(\mathbf{x})}\mathbf{x} - \mathbf{H}^T\mathbf{p}_m,
		\end{equation}
		where 
		\begin{equation}
			\label{eq:axdef}
			\mathbf{A}^{(\mathbf{x})} =  
			\textbf{H}^{T}\textbf{H} +
			\lambda \alpha \mathbf{W}^{(\mathbf{x})} + 
			\lambda (1-\alpha)\sum_{i} \mathbf{D}_{o,i}^{T}\mathbf{W}^{(\mathbf{x})} \mathbf{D}_{o,i} 
			+  \lambda_{p}\; \mathbf{N}^{(\mathbf{x})} .
		\end{equation}
		Here  $\mathbf{W}^{(\mathbf{x})}$ and $\mathbf{N}^{(\mathbf{x})}$  are diagonal matrices with $j$th diagonal element given by
		\begin{align}
			\label{eq:wx}
			\{\mathbf{W}^{(\mathbf{x})}\}_{jj}  & = q\left(\epsilon + \alpha (\mathbf{x})_{j}^{2} +
			(1-\alpha) \sum_{i}\big(\textbf{D}_{o,i}\mathbf{x}\big)_{j}^{2}\right)^{q-1},  \\
			\label{eq:nx}
			\{\mathbf{N}^{(\mathbf{x})}\}_{jj} & = 0.5(1 - sign((\mathbf{x})_j)).
		\end{align}
		The superscript  `$(\mathbf{x})$'  in the diagonal matrices signify their dependence on $\mathbf{x}$, which makes the operation    $\mathbf{A}^{(\mathbf{x})}\mathbf{x}$ a non-linear operation. 
		
		The proposed preconditioned gradient search proceeds as follows:  given a current estimate of the
		minimum,  $\mathbf{x}^{(k)}$,   we update the estimate as  
		$\mathbf{x}^{(k+1)}= \mathbf{x}^{(k)}-
		\beta_k \mathbb{P}(\nabla J({\bf x}^{(k)}, q))$, where 
		$\mathbb{P}(\cdot)$ is the preconditioning, and  $\beta_k$ is the step-size.
		To describe how the preconditioning is done,
		let $\mathbf{g}^{(k)}=  \nabla J({\bf x}^{(k)}, q)$  and  
		$\hat{\mathbf{g}}^{(k)} =  \mathbb{P}(\nabla J({\bf x}^{(k)}, q))$.  Then 
		$\hat{\mathbf{g}}^{(k)}$  is determined by solving the linear system of equations
		$\mathbf{A}^{(\mathbf{x}^{(k)})}\hat{\mathbf{g}}^{(k)}= {\mathbf{g}}^{(k)}$, where
		\begin{equation}
			\label{eq:pcond}
			\mathbf{A}^{(\mathbf{x}^{(k)})} = \mathbf{H}^{T}\mathbf{H} +
			\lambda  \alpha \mathbf{W}^{(\mathbf{x}^{(k)})} + \lambda (1-\alpha)
			\sum_{i} \mathbf{D}_{o,i}^{T}\mathbf{W}^{(\mathbf{x}^{(k)})} \mathbf{D}_{0,i} 
			+ \lambda_{p}\; \mathbf{N}^{(\mathbf{x}^{(k)})} .
		\end{equation}
		The idea behind this preconditioning is to mimic 
		damped Newton approach as done in \cite{nestedcg}. Here the matrix $\mathbf{A}^{(\mathbf{x}^{(k)})}$, which is the matrix
		involved in the gradient expression  of equations \eqref{eq:graddef}, \eqref{eq:axdef},
		works as an   approximation for Hessian of $J(\cdot, q)$. To solve the equation
		$\mathbf{A}^{(\mathbf{x}^{(k)})}\hat{\mathbf{g}}^{(k)}= {\mathbf{g}}^{(k)}$, 
		we use the well-known method of conjugate gradients (CG)  and
		denote the operation
		by $\hat{\mathbf{g}}^{(k)} = CG(\mathbf{A}^{(\mathbf{x}^{(k)})}, {\mathbf{g}}^{(k)}, \epsilon_{cg})$
		where $\epsilon_{cg}$ denotes the  termination tolerance for CG iteration.
		Next, the step size  $\beta_k$ is  chosen such
		that $\frac{J({\bf x}^{(k+1)},q)}{J({\bf x}^{(k)},q)} < \epsilon_{s}$ by means of a  back-tracking procedure,
		where $\epsilon_{s}$ is an another user-defined tolerance for back-tracking.
		Specifically,  starting with $\beta_k=1$,  the required $\beta_k$ is determined
		by series of   checks on the condition $\frac{J({\bf x}^{(k+1)}, q)}{J({\bf x}^{(k)}, q)} < \epsilon_{s}$
		with iterated multiplication of  $\beta_k$ with a factor $\rho \in (0,1)$.   Finally, the iterative
		update on $\mathbf{x}^{(k)}$ is terminated upon the attainment of the condition
		$\cfrac{ ||\mathbf{x}^{(k+1)} - \mathbf{x}^{(k)} ||_{2}  }{||  \mathbf{x}^{(k)} ||_{2}} < \epsilon_o$
		where $\epsilon_o$ is  a yet another user-defined tolerance.  This iterative method is summarized
		in the panel {\bf Algorithm  \ref{alg:RR}},  where the first input $\mathbf{x}^{(0)}$ is an appropriate initialization.
		{\bf Algorithm  \ref{alg:RR}} calls the back-tracking line search method described above.  This line search is summarized
		in the panel  {\bf Algorithm  \ref{alg:LS}}.

		\begin{algorithm}[ht]
			\DontPrintSemicolon
			\SetAlgoLined
			$\textbf{Initialization}: \;   k \gets 0, \;   r_k \gets 1+\epsilon_o$ \\
			\While{$r_k  \geq \epsilon_o$} 
			{
				$ \mathbf{g}^{(k)} \gets \nabla J(\mathbf{x}^{(k)}, q)$  \Comment{Use Eqs. \eqref{eq:graddef},  \eqref{eq:axdef}, \eqref{eq:wx}, and \eqref{eq:nx} }\\
				$\hat{\mathbf{g}}^{(k)} \gets \text{CG}(\mathbf{A}^{(\mathbf{x}^{(k)})},\mathbf{g}^{(k)}, \epsilon_{cg})$ 
				\Comment{Use Eq.  \eqref{eq:axdef}. CG:  Conjugate gradient method} \\
				$\beta_{k} \gets \text{LS}(\textbf{x}^{(k)},\hat{\mathbf{g}}^{(k)}, \rho, \epsilon_s)$ \Comment{Line search ({\bf Algorithm \ref{alg:LS}})}\\
				$\textbf{x}^{(k+1)} \gets \textbf{x}^{(k)} - \beta_{k}  \hat{\mathbf{g}}^{(k)}$\\
				$r_{k+1} \gets \cfrac{ ||\mathbf{x}^{(k+1)} - \mathbf{x}^{(k)} ||_{2}  }{||  \mathbf{x}^{(k)} ||_{2} }$\\
				$k \gets k+1 $ \\
			}
			\Return $\mathbf{y} = \textbf{x}^{(k)}$
			\caption{Regularized Reconstruction Algorithm:  
				$RR(\mathbf{x}^{(0)}, \; \mathbf{p}_m, \;
				\alpha,\; \lambda,\; \lambda_{p},\;    \; \epsilon_s, \;   \epsilon_{cg},\; \epsilon_{o},  \rho, q)$}
			\label{alg:RR}
		\end{algorithm}

		\begin{algorithm}[H]
			\DontPrintSemicolon
			\SetAlgoLined
			$\textbf{Initialization}:\;  \hat{\beta} \gets 1,  \;  D \gets J(\textbf{x}^{(k)}), \;   
			N \gets J(\textbf{x}^{(k)}-\hat{\beta}\hat{\mathbf{g}}^{(k)})$     \Comment Use Eq. \eqref{eq:ucprob} \\
			\While{$\frac{N}{D} \ge  \epsilon_s$}
			{
				$\hat{\beta} \gets \rho \hat{\beta} $ 
				
				$N \gets J(\textbf{x}^{(k)}-\hat{\beta}\hat{\mathbf{g}}^{(k)})$   \Comment Use Eq. \eqref{eq:ucprob} 
				
			}
			\Return $\beta_{k}=\hat{\beta}$
			\caption{Line Search Algorithm:
				$ \beta_k = \text{LS}(\textbf{x}^{(k)},\hat{\mathbf{g}}^{(k)}, \rho, \epsilon_s)$} 
			\label{alg:LS}
		\end{algorithm}
		
		\begin{algorithm}[H]
			\DontPrintSemicolon
			\SetAlgoLined
			
			$\textbf{Initialization}:  
			\bar{\bf A} \gets \textbf{H}^{T}\textbf{H} + 
			\lambda\alpha {\bf I} + \lambda (1-\alpha)\sum_{i}\textbf{D}_{o,i}^{T}\textbf{D}_{o,i}$\\
			$~~~~~~~~~~~~~~~~~~~~~~~\mathbf{y}^{(-1)} \gets CG(\bar{\bf A}, \textbf{H}^{T} \textbf{p}_m,
			\epsilon_{cg})$\\
			\For {$m =0,1,\ldots, n_s$}
			{
				$q_m \gets 0.5 - m\frac{0.5-q}{n_s}$ 
				
				$\mathbf{y}^{(m)} = RR( \mathbf{y}^{(m-1)}, \; \mathbf{p}_m, \;
				\alpha,\; \lambda,\; \lambda_{p},  \; \epsilon_s, \;   \epsilon_{cg},\; \epsilon_o, \;  \rho, q_m)$ \Comment{\bf Algorithm \ref{alg:RR}}

				$m \gets m+1$ 
				
			}
			\Return $\mathbf{y}^*=\mathbf{y}^{(n_s)}$
			\caption{Graduated Non-Convexity Algorithm:  GNC( $\mathbf{p}_m, \;
				\alpha,\; \lambda,\; \lambda_{p},\;   \; \epsilon_s, \;   \epsilon_{cg},\;  \epsilon_{o}, \rho, q, n_s$)
			} 
			\label{alg:GNC}
		\end{algorithm}

		Since we choose $q$  to be less than $0.5$, the regularization functional is non-convex. 
		Hence the cost function can have several local minima, and the tolerances
		used in the inner iterations ($\epsilon_{cg}$ and $\epsilon_s$) might influence the actual
		minimum attained by the overall algorithm.  Such a minimum is likely to contain artifacts.
		To alleviate this problem,  we adopt the well-known
		graduated non-convexity (GNC)  approach \cite{nikolova2010}.
		Specifically, we introduce an outer loop in reconstruction method,
		where  a series of $n_s$ regularized reconstructions are performed with sparsity indices 
		$\{q_m:=0.5-m(0.5-q)/n_s, m = 0,1, \ldots, n_s\}$. For each sparsity index $q_m$,  the 
		reconstruction is denoted by $\mathbf{y}^{(m)}$, and it is obtained by calling regularized 
		reconstruction ({\bf Algorithm \ref{alg:RR}})   with the initialization set to  $\mathbf{y}^{(m-1)}$,
		which is the reconstruction corresponding to the previous sparsity index $q_{m-1}$.  Note that, for $m=0$, 
		$q_0= 0.5$,  and for this value of $q_0$, the cost is convex and hence the parameters
		$\epsilon_{cg}$ and $\epsilon_s$  do not affect the reconstruction obtained ($\mathbf{y}^{(0)}$).  
		For subsequent calls for
		reconstruction
		with iteration index  $m > 0$,  since   the sparsity index $q_m$ is typically close to the index of previous
		reconstruction,  $q_{m-1}$,   the reconstruction $\mathbf{y}^{(m)}$ is likely to be close to $\mathbf{y}^{(m-1)}$.
		As a result,  the sequence $\{\mathbf{y}^{(m)}, m=0,1,\ldots, n_s\}$ is mostly determined by the sequence
		$\{q_m, m = 0,1, \ldots, n_s\}$.  This will  
		reduce the dependence of the final 
		reconstruction on the parameters  $\epsilon_{cg}$ and $\epsilon_s$  and will  lead a good
		quality final reconstruction.
		For  $m=0$, we need an efficient
		initialization for calling the regularized reconstruction.
		We propose to use the result of  minimizing $J({\bf \cdot}, q)$ with $q=1$. 
		With $q=1$,  $J({\bf \cdot}, q)$ is quadratic  and its minimum is determined
		by the following linear system of equations:
		
		\begin{equation}
			\big[ \textbf{H}^{T}\textbf{H} + 
			\lambda\alpha {\bf I} + \lambda (1-\alpha)\sum_{i}\textbf{D}_{o,i}^{T}
			\textbf{D}_{o,i}\big]\mathbf{y}^{(-1)} =   \textbf{H}^{T} \textbf{p}_m.
		\end{equation} 
		We solve this again by calling the conjugate gradient method.
		The overall algorithm is given in the {\bf Algorithm \ref{alg:GNC}}. 
		Note that the possibility of using the GNC method is the main advantage of replacing the
		logarithm used in the original formulation  \cite{arigovindan2013high} by the fractional power $q$.

		Now, we will specify the modifications
		required if $R_{h,2}$ is used.  It is only required to change the operator 
		$\mathbf{A}^{(\mathbf{x})}$ present in the equation (\ref{eq:axdef}).  The operator
		corresponding to the
		regularization $R_{h,2}$ can be expressed as
		
		\begin{equation}
			\label{eq:axdef2}
			\mathbf{A}^{(\mathbf{x})} =  
			\textbf{H}^{T}\textbf{H} +
			\lambda \alpha \bar{\mathbf{W}}^{(\mathbf{x})} +  
			\lambda (1-\alpha) \sum_{i} \mathbf{D}_{o,i}^{T}\hat{\mathbf{W}}^{(\mathbf{x})} \mathbf{D}_{o,i} 
			+  \lambda_{p}\; \mathbf{N}^{(\mathbf{x})} .
		\end{equation}
		where $\bar{\mathbf{W}}^{(\mathbf{x})}$  and $\hat{\mathbf{W}}^{(\mathbf{x})}$  are diagonal matrices 
		with its elements given by
		\begin{align}
			\{\bar{\mathbf{W}}^{(\mathbf{x})}\}_{jj} & =  q\left(\epsilon + (\mathbf{x})_{j}^{2} \right)^{q-1}, 
		\end{align}
		\begin{align}
			\{\hat{\mathbf{W}}^{(\mathbf{x})}\}_{jj} & =  q\left(\epsilon  +  \sum_{i}\big(\textbf{D}_{o,i}\mathbf{x}\big)_{j}^{2}\right)^{q-1}.
		\end{align}
		
		\subsection {Matrix-free implementation of the  forward model}
		
		Finding the gradient $\mathbf{g}^{(k)}$ of the cost function and the search direction 
		$\hat{\mathbf{g}}^{(k)}$
		using CG involves the  repeated application of $\textbf{H}^{T}\textbf{H} $; this is computationally 
		very expensive and needs large amount of memory. For example,  generating $\textbf{H}$ 
		for  image size  $512 \times 512$ corresponding to  $128$  transducers with 
		each transducer taking $1024$  samples results in $10^5$ elements in $\mathbf{H}$, and
		thus requires 
		256 GB RAM \cite{rosenthal2010fast}.  In this work, we derive a formula for matrix free implementation 
		of the proposed algorithm by doing on-the-fly computation of  multiplication with 
		$\textbf{H}^{T}\textbf{H}$ without explicitly constructing $\mathbf{H}$. 
		
		We first note that $\mathbf{H}$ is of the form  
		$\mathbf{H} = [\mathbf{H}_1^T\;\mathbf{H}_2^T\; \cdots \mathbf{H}_M^T]^T$,  where 
		$\mathbf{H}_i$  represents the operation of obtaining samples from $L$ transducers at $i$th
		time instant.   Let $\mathbf{v}$  be  some vector that undergo multiplication by $\mathbf{H}$.
		Let $v(\mathbf{r})$ be the image obtained by putting elements of $\mathbf{v}$ into image form.
		Then the operator  that is the equivalent of multiplying $\mathbf{v}$  with 
		$\mathbf{H}_i$ can be written by using the forward model of the equation
		\eqref{eq:coxexp}  as
		\begin{equation}
			\mathbf{H}_i \mathbf{v} = 
			\mathcal{P}\mathcal{F}^{-1}
			\left\{ \mathcal{F}
			(\mathcal{A}\mathbf{v})
			\cos (c_{0}|\textbf{k}|t_i)\right\},
		\end{equation}
		where  $\mathcal{P}$ is the operation that represents retrieval  of samples from 
		transducer locations  $\{\mathbf{r}_s,s=1,\ldots,L\}$ into vector form,  and 
		$\mathcal{A}$ represents the operation of assembling the $N\times 1$ vector 
		into an $N_x\times N_y$ image.  
		Then, multiplication with    $\mathbf{H}_i^T$  can be expressed as
		\begin{equation}
			\mathbf{H}_i^T\mathbf{u}  =   
			\mathcal{A}^a
			\mathcal{F}^{-1} \left\{
			\cos (c_{0}|\textbf{k}|t_i)
			\mathcal{F}
			\left\{
			\mathcal{P}^a\mathbf{u}
			\right\}
			\right\},
		\end{equation}
		where  $\mathcal{A}^a$  and  $\mathcal{P}^a$ are the adjoints of   $\mathcal{A}$  and  $\mathcal{P}$
		respectively.      $\mathcal{A}^a$   represents scanning a $N_x\times N_y$  image into a vector of
		size $N\times 1$ ($N=N_xN_y$).    Further,   assuming that the transducer locations 
		$\{\mathbf{r}_s,s=1,\ldots,L\}$  are  in a subset of image grid points,  
		$\mathcal{P}^a$      becomes the operation of embedding
		an $L\times 1$ vector  into an  zero-image of size   $N_x\times N_y$. 
		Then,  the operator-equivalent of  $\mathbf{H}_i^T\mathbf{H}_i$  can be expressed  as
		\begin{equation}
			\mathbf{H}_i^T\mathbf{H}_i \mathbf{v}  =   
			\mathcal{A}^a
			\mathcal{F}^{-1}
			\left\{ \cos (c_{0}|\textbf{k}|t_i)
			\mathcal{F}\left\{
			\mathcal{P}^a\mathcal{P}\mathcal{F}^{-1}
			\left\{ \mathcal{F}(\mathcal{A}\mathbf{v})\cos (c_{0}|\textbf{k}|t_i)\right\} \right\} \right\}
		\end{equation}
		It can be shown that $\mathcal{P}^a\mathcal{P}$ becomes equivalent to
		multiplication by a binary image of $1$s and $0$s,  with $1$s corresponding to the transducer locations.
		Let $S(\mathbf{r})$ be this binary image.
		
		\begin{equation}
			\mathbf{H}_i^T\mathbf{H}_i \mathbf{v}  =   
			\mathcal{A}^a
			\mathcal{F}^{-1}
			\left\{ \cos (c_{0}|\textbf{k}|t_i)
			\mathcal{F}\left\{
			S(\mathbf{r})\mathcal{F}^{-1}
			\left\{ \mathcal{F}(\mathcal{A}\mathbf{v})\cos (c_{0}|\textbf{k}|t_i)\right\} \right\} \right\}
		\end{equation}
		This results in the following form for 
		$\mathbf{y} = \mathbf{H}^T\mathbf{H} \mathbf{v}$
		\begin{align}
			\mathbf{y} &= \mathbf{H}^T\mathbf{H} \mathbf{v}  = \sum_{i=1}^M   \mathbf{H}_i^T\mathbf{H}_i \mathbf{v}  \\
			& =
			\mathcal{A}^a
			\sum_{i=1}^M 
			\mathcal{F}^{-1}
			\left\{ \cos (c_{0}|\textbf{k}|t_i)
			\mathcal{F}\left\{
			S(\mathbf{r})\mathcal{F}^{-1}
			\left\{ \mathcal{F}(\mathcal{A}\mathbf{v})\cos (c_{0}|\textbf{k}|t_i)\right\} \right\} \right\}
		\end{align}
		
		In actual implementation,  we do not build vectors.  We used vector-matrix notation in the description given above only for notational convenience.  Every vector involved in the
		algorithm is kept in the image form.  Specifically,  the above operator's input will be in image form, and the output will also be in image form.   If $v(\mathbf{r})$ is the $N\times N$ image equivalent of 
		the $N^2\times 1$ vector $\mathbf{v}$ ,  and $y(\mathbf{r})$ is the image equivalent of the vector $\mathbf{y}$
		the above operator can be represented by
		
		\begin{equation}
			\label{eq:hthfourier}
			y(\mathbf{r})
			=  \sum_{i=1}^M 
			\mathcal{F}^{-1}
			\left\{ \cos (c_{0}|\textbf{k}|t_i)
			\mathcal{F}\left\{
			S(\mathbf{r})\mathcal{F}^{-1}
			\left\{ \mathcal{F}(v(\mathbf{r}))\cos (c_{0}|\textbf{k}|t_i)\right\} \right\} \right\}
		\end{equation}

		\section{Reconstruction results}

		\subsection{Reconstruction from simulated data}
		
\begin{figure}[http]
			\centering
			\fbox{\includegraphics[width= 0.7\linewidth]{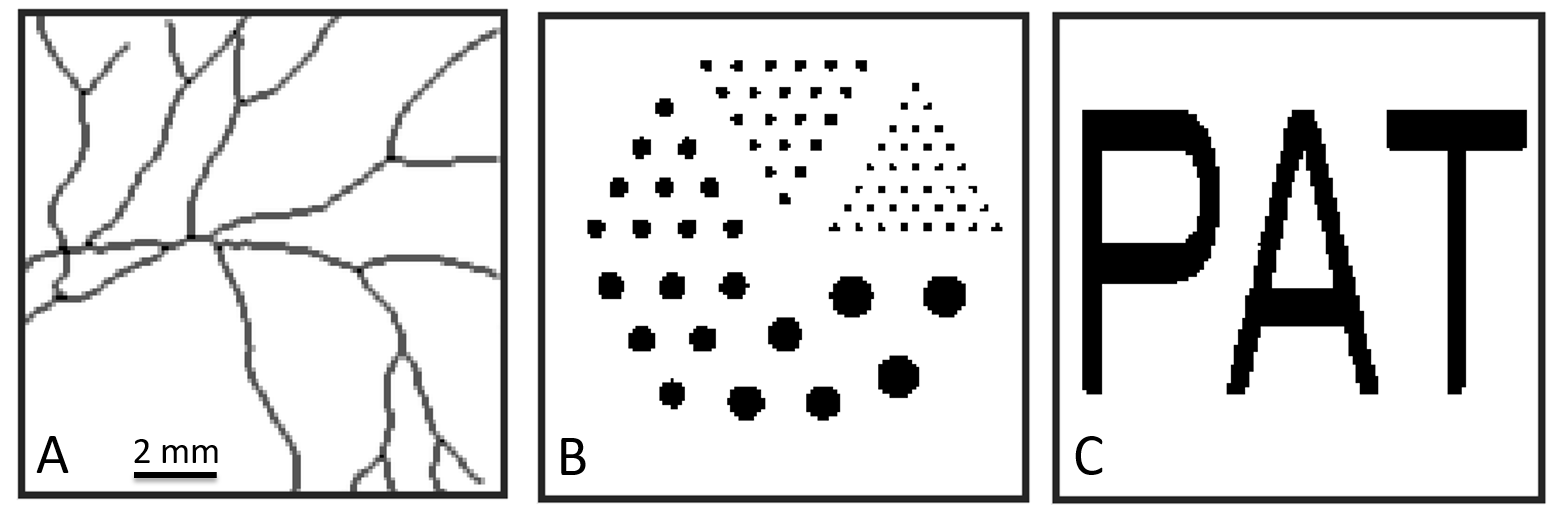}}
			\caption{Numerical phantoms used for  evaluation of the proposed method}
			\label{fig:phantomlist}
		\end{figure}

\begin{figure}[http]
			\centering
			\fbox{\includegraphics[width= 0.4\linewidth]{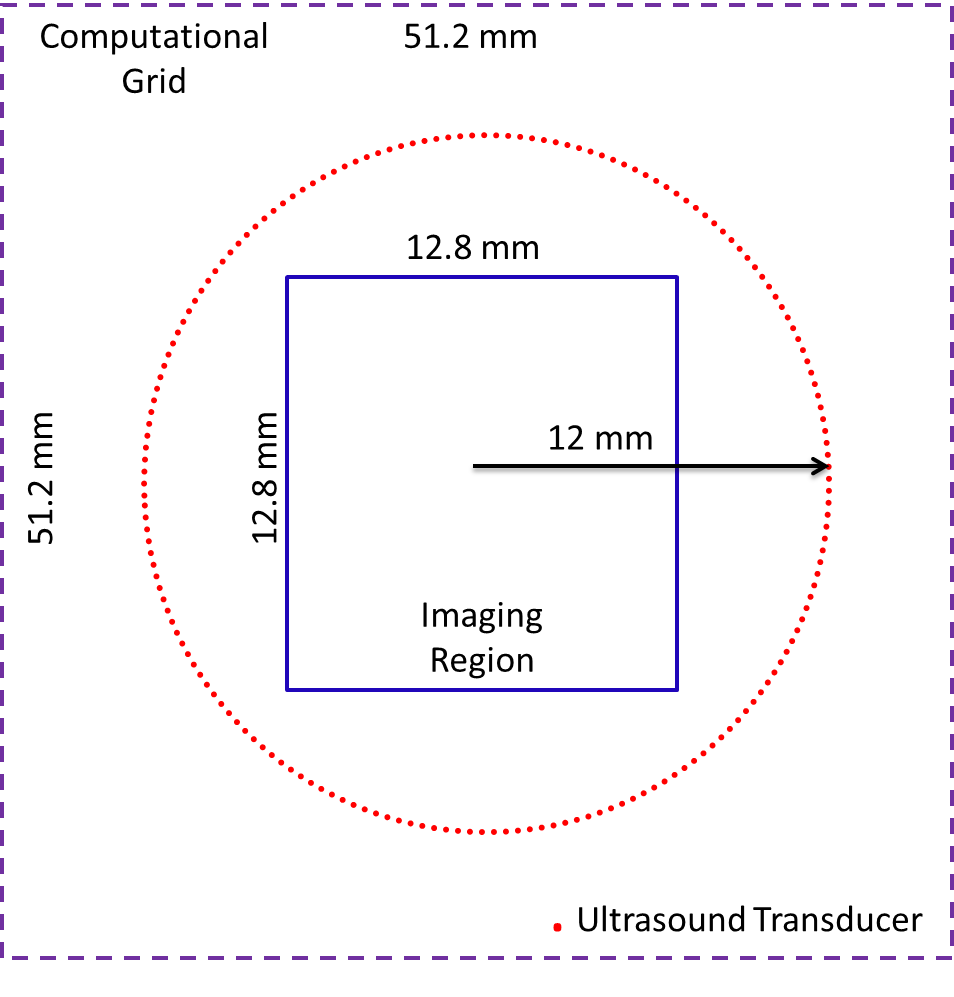}}
			\caption{a) Schematic diagram of PA data acquisition geometry with ultrasound transducers 
				(shown by dots) around the imaging region of 12.8 mm $\times$ 12.8 mm. The computational imaging grid size is   51.2 mm $\times$  51.2 mm.}
			\label{fig:synthscheme}
		\end{figure}

		We used three numerical phantoms that are commonly used to evaluate PAT reconstruction methods viz. Blood vessel,  Derenzo,   and  PAT, which is given in Figure \ref{fig:phantomlist}.
		All were normalized to the size $128\times 128$   with a corresponding physical size of 12.8 mm $\times$ 12.8 mm as a model for generating the synthetic data.
		The synthetic data was generated as per the geometry given in Figure \ref{fig:synthscheme}, where the dotted circle, whose radius is 12 mm,  denotes the  trajectory of possible locations for 
		transducers. The required image (image to be reconstructed) was defined to be on a larger grid of size $512\times 512$ with the equivalent physical size of  51.2 mm $\times$ 51.2 mm.  
		This was done to accommodate for the boundary effect caused
		by  Fourier based convolutions  represented in Equation \eqref{eq:hthfourier}.
		The forward data was generated using  Equation \eqref{eq:coxexp} and added with Gaussian noise to form simulated data having  SNR levels of 20 dB, 30 dB, and 40 dB.  Values for $t$ in the model
		of  Eq. \eqref{eq:coxexp}  was chosen  as $\{\delta_ti, i=1,\ldots,M\}$ with $1/\delta_t=100$ MHz and
		$M=1600$. The number of transducers to mimic the limited data scenario in our experiments was taken as 16, 32, 64, and 128. 
		Note that most reported methods in the literature typically use not less than 128 transducers. The sound speed in the medium was assumed to be $1.5 \; mm/\mu s$, and we considered the medium to be homogeneous with no dispersion or absorption of sound.  We choose second-order derivatives for the regularization, i.e.,  we set 
		$o=2$ in the equation \eqref{eq:rh1} and \eqref{eq:rh2}. The parameter $\lambda$  was determined using the model itself, as done in most methods that focus on the design of regularization.  Next, for positivity,  we set $\lambda_p = 10\lambda$. Further, 
		we found that $q=0.25$ with $n_s=10$ was adequate for all cases. All the tolerances ( $\epsilon, \epsilon_s, \epsilon_{cg}$, and $\epsilon_{o}$)  were set to $10^{-6}$. Further, for all test cases, setting $\alpha=0.5$ gave good quality reconstructions. 
		For the blood vessel phantom,  we conducted experiments with the number of transducers set to 64 and 128. For the remaining phantoms,  we did experiments with 16 and 32 as the number of transducers.  For a given $q$,  each reconstruction takes about 38 mins for the proposed method in Dell Precision T7820, whereas the FISTA-based methods take about 30 mins.
		
		
		Considering the choice of method to compare, we first note that
		derivative based regularization methods are known to deliver 
		robust and consistent performance in most inverse problems
		and hence we restrict to  this class for choosing the method
		to compare.  These methods can be categorized into 
		Tikhonov filtering methods 	\cite{buehler2011model} and
		total variation methods
		\cite{ 	arridge2016accelerated,	huang2013full,
			arridge2016adjoint,
			boink2018framework,
			han2015sparsity, 
			wang2012investigation}, and the later ones are more
		robust. 
		Among the total variation methods,
		the method of Boink et al \cite{boink2018framework} use 
		total generalized variation (TGV) and others use first
		order total variation (TV-1).  Although TGV is known to be superior
		in its formulation,  it poses  numerical difficulties
		in implementing the reconstruction. This narrows
		down our choice of method to compare to be within TV-1 methods.
		All TV-1  methods  	\cite{ arridge2016accelerated,	huang2013full,
			arridge2016adjoint,
			han2015sparsity, 
			wang2012investigation}
		differ only by speed and we chose the method of Huang et al 
		\cite{huang2013full}  because of its ease of implementation.
		We compare the proposed
		method using {\em structural similarity (SSIM) index} \cite{ssim}
		as it is more sensitive to reconstruction artifacts.

\begin{figure}[H]
			\centering
			\fbox{\includegraphics[width=0.7\linewidth]{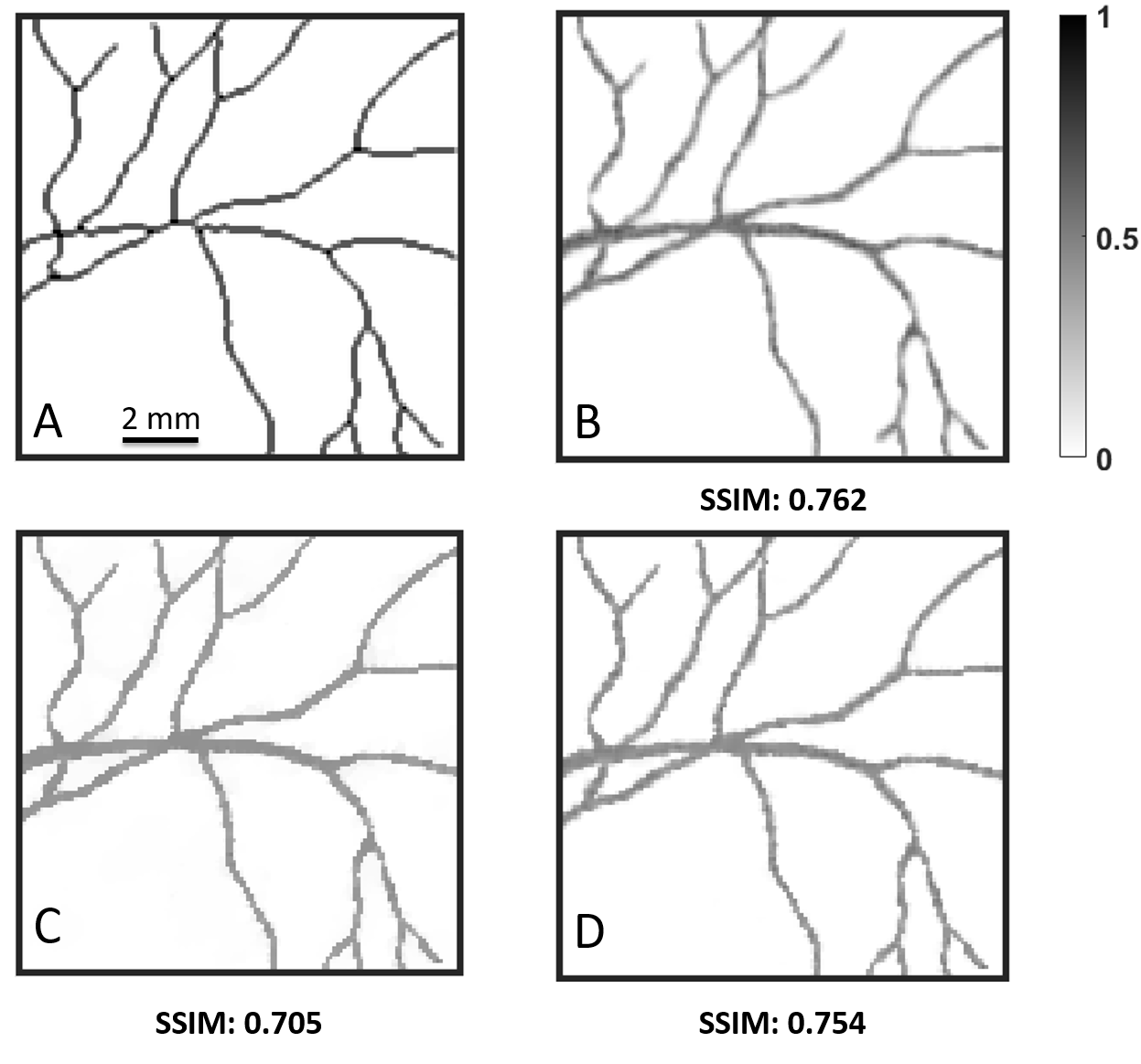}}
			\caption{Comparison of reconstructions obtained from simulated data with 128 transducers and 20 dB measurement noise.  
				(A): reference phantom model (the maximum initial Pressure rise 
				is assumed to be 1Pa);  
				(B):  reconstruction obtained by  proposed method with form I regularization ;
				(C) reconstruction from FISTA-based method  with $\lambda$ chosen for best SSIM 
				score ; 
				(D): reconstruction obtained by  proposed method with form II regularization. 
			}
			\label{fig:reccomp1}
		\end{figure}

\begin{figure}[H]
			\centering
			\fbox{\includegraphics[width=0.65\linewidth]{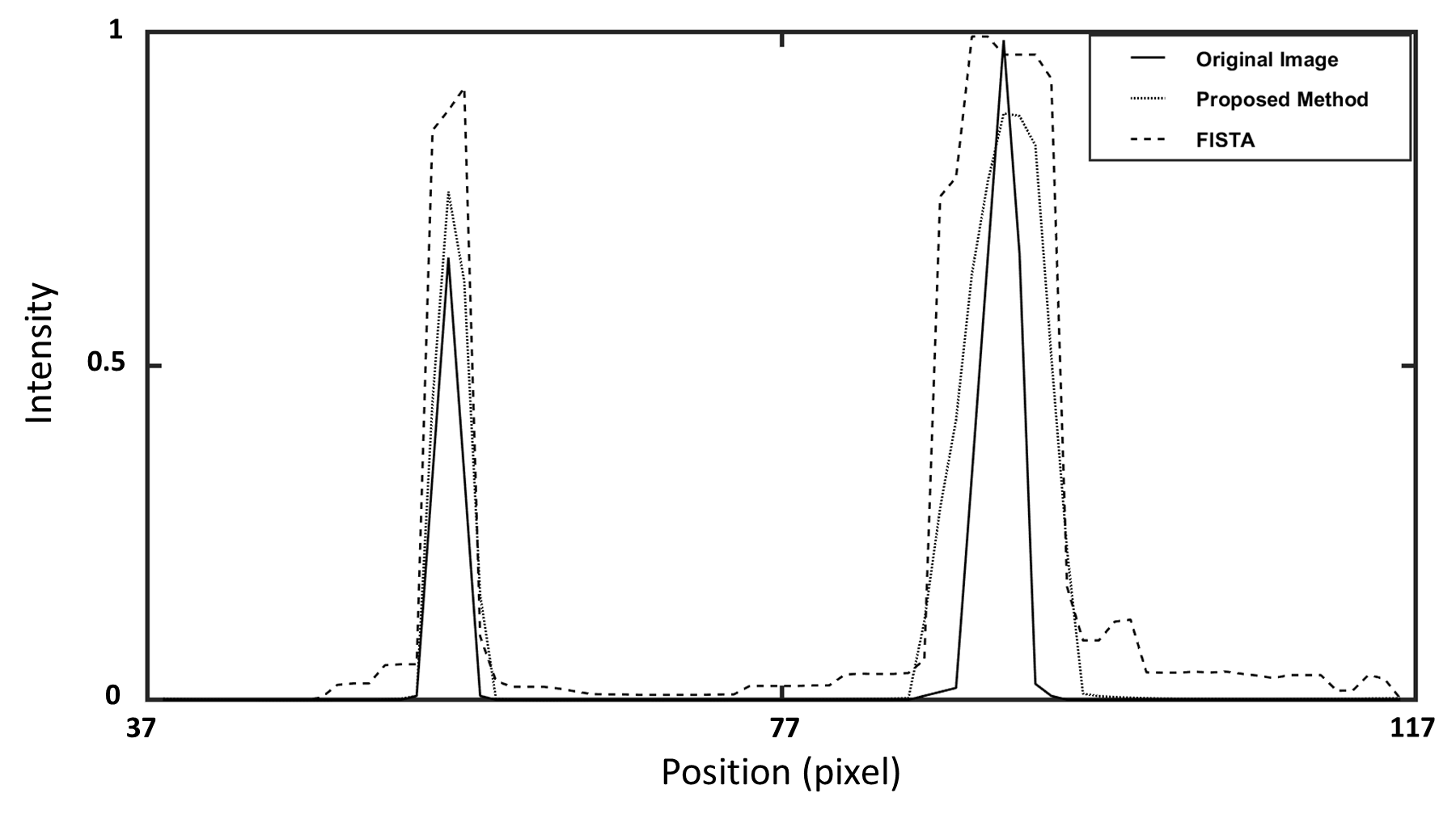}}
			\caption{Scan line based intensity profiles of reconstructed images from Fig:\ref{fig:reccomp1}.  
			}
			\label{fig:fig2scan}
		\end{figure}

		In the first experiment,  we generated measurements corresponding to 128 transducers with an SNR of  20 dB. Figure \ref{fig:reccomp1}  compares the reconstructed results obtained from this measurement set.  The proposed method with both regularization forms yields better reconstruction over the FISTA-based method, although the result with the second form of regularization gives a slightly inferior result.  This is reflected by the relative improvement in the  SSIM score (0.057 and 0.049). The proposed method was also able to reconstruct the peak amplitude of the initial pressure  distribution (1 Pa) more accurately than the FISTA algorithm as  evident from the  Figures \ref{fig:reccomp1}.A,   \ref{fig:reccomp1}.B,    \ref{fig:reccomp1}.C and \ref{fig:reccomp1}.D.   Figure \ref{fig:fig2scan} shows the scanline based intensity profile of reconstructions in Figure \ref{fig:reccomp1}. The intensity profiles show that our method closely follows the ground truth compared to the FISTA based method, which is also confirmed by superior SSIM scores.
		Next,  we consider two more noise levels viz. 30 dB and 40 dB.  
		For this phantom, since the first form of proposed regularization outperforms the second, we present the reconstruction results only for the first form for these additional input SNR cases.
		The relative improvement in our SSIM scores over the FISTA-based method corresponding to 30 dB and 40 dB are 0.026 and 0.022, respectively.  The proposed method yielded an SSIM score of   0.7622 for a 20 dB noise level, whereas the FISTA-based produced 0.7052. This indicates that the proposed method retrieves most of the quality that can be obtained from nearly noise-free data from the noisy data.  On the other hand, the improvement in reconstruction yielded by the FISTA-based method concerning an increase in the input SNR is more gradual, as indicated by the SSIM scores  0.7052,  0.7370, and  0.7427 corresponding to input SNRs 20 dB, 30 dB, and 40 dB.  Further,  it also clear that the difference in performance between the proposed and FISTA-based methods is higher for lower input SNR.    Next, we repeated the above trials with 64 transducers instead of 128. The relative improvements in the proposed method's SSIM scores with the first regularization form are 0.118, 0.039, and 0.019 for the corresponding input SNRs 20 dB, 30 dB, and 40 dB.

		\begin{figure}[ht]
			\centering
			\fbox{\includegraphics[width=0.8\linewidth]{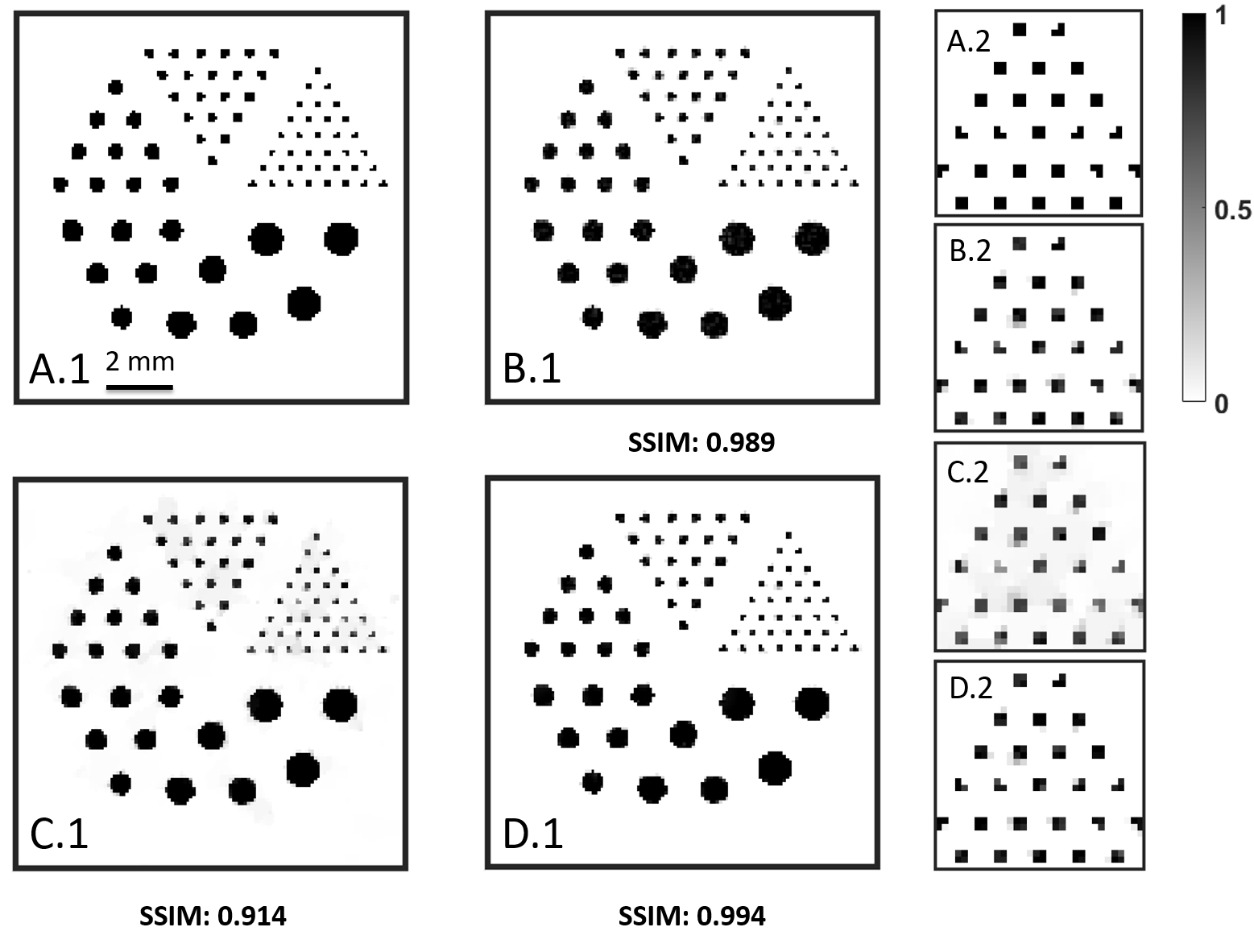}}
			\caption{Comparison of reconstructions obtained from simulated data with 
				32 transducers and 20 dB measurement noise.  
				(A.1): Derenzo phantom model (the maximum initial Pressure rise 
				is assumed to be 1Pa);  
				(B.1):  reconstruction obtained by  proposed method with form I regularization ;
				(C.1) reconstruction from FISTA-based method  with $\lambda$ chosen for best SSIM 
				score; 
				(D.1): reconstruction obtained by  proposed method with form II regularization. A.2, B.2, C.2, and D.2  are cropped regions from 
				A.1, B.1, C.1, and D.1.}
			\label{fig:reccomp5}
		\end{figure}
		In the next experiment,  we generated measurements using a Derenzo phantom image corresponding to 32 transducers with an SNR of  20 dB. Figure \ref{fig:reccomp5}  compares the reconstructed results obtained from this measurement set. Here too, the proposed method with both forms of regularization yields a better reconstruction over the FISTA-based method as seen from the  Figures \ref{fig:reccomp5}.A.1,   \ref{fig:reccomp5}.B.1,    \ref{fig:reccomp5}.C.1, and \ref{fig:reccomp5}.D.1.   Figures
		\ref{fig:reccomp5}.A.2,   \ref{fig:reccomp5}.B.2, \ref{fig:reccomp5}.C.2, and \ref{fig:reccomp5}.D.2  display a cropped region from the images displayed in \ref{fig:reccomp5}.A.1,   \ref{fig:reccomp5}.B.1,    \ref{fig:reccomp5}.C.1, and \ref{fig:reccomp5}.D.1 for a closer view.  The improvement seen in the displayed images is also reflected by the relative improvement in the SSIM score (0.074 and 0.08). Next, we considered two other noise levels, viz.,  30 dB, and 40 dB, for the same phantom with the same number of transducers (32).   For these cases,  we present the reconstruction result only for the second form since the second form of proposed regularization outperforms the first for the current phantom. The relative improvement in the SSIM scores of the reconstructions from data sets with input SNRs 30 dB and 40 dB is 0.017 and 0.01. This again proves that the proposed method retrieves most of the quality obtained from nearly noise-free data from the noisy data. The performance between the proposed and FISTA is higher for lower input SNR in this experiment also.
		\begin{figure}[ht]
			\centering
			\fbox{\includegraphics[width=0.8\linewidth]{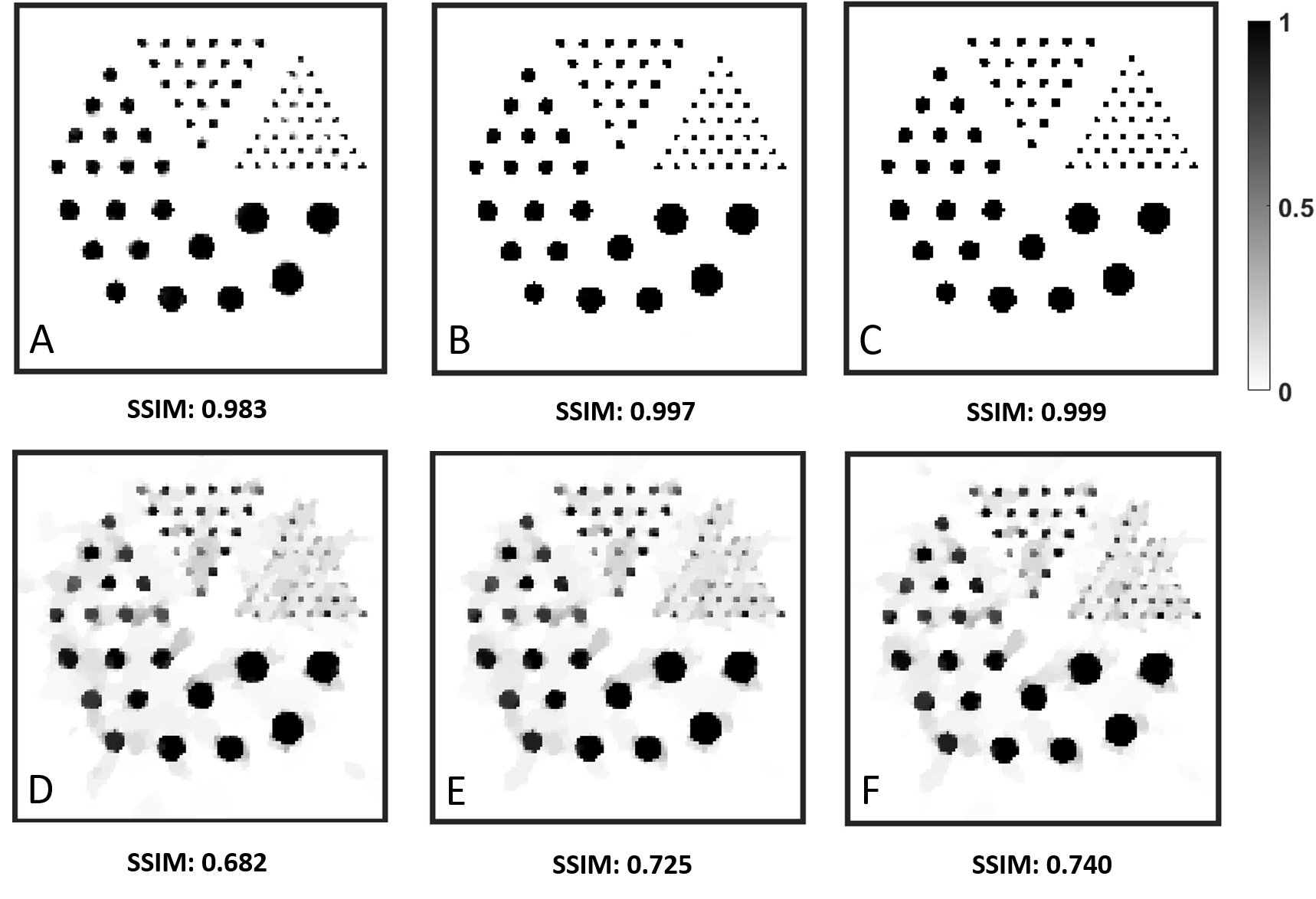}}
			\caption{Reconstructed images from input datasets corresponding to 16 transducers for various noise levels. (A), (B), and (C):   reconstructions obtained from the proposed method with second form of regularization  corresponding to SNRs 20 dB, 30 dB, and 40 dB.
				(D),  (E), and  (F): reconstructions obtained from the FISTA method corresponding to SNRs 20 dB, 30 dB, and 40 dB.}
			\label{fig:reccomp7}
		\end{figure}
		Next, we repeated the same set of trials with 16 transducers instead of 32. The proposed method outperformed the FISTA method for all SNRs levels, in this case, also as shown in Figure  \ref{fig:reccomp7}.
		The quality of the reconstruction for the FISTA method was heavily degraded for all SNRs (20 dB, 30 dB, and 40 dB), and the reconstructed images contained artifacts, as seen from Figure  \ref{fig:reccomp7}.D, Figure  \ref{fig:reccomp7}.E and Figure  \ref{fig:reccomp7}.F respectively. However, the reconstruction quality was not much affected by the proposed method, although the number of transducers used is very less. The proposed method's improved performance in comparison with the FISTA method is also reflected in the large difference of SSIM scores (0.301, 0.272, and 0.259) for all three input SNR levels: 20 dB, 30 dB, and 40 dB, respectively. These results demonstrate the robustness of our algorithm in the limited data scenario. The above observations were further verified by repeating the same pattern of experiments on the PAT phantom. Reconstruction results for this phantom and the results of all other trials reported in this section are displayed  {\bf Table \ref{table:1}}. For a visual comparison of performance evaluated using PAT phantom, we display the images corresponding to trials with 16 transducers in Figure \ref{fig:reccomp10}, which confirms the superiority of the proposed method.

		\begin{figure}[H]
			\centering
			\fbox{\includegraphics[width=0.8\linewidth]{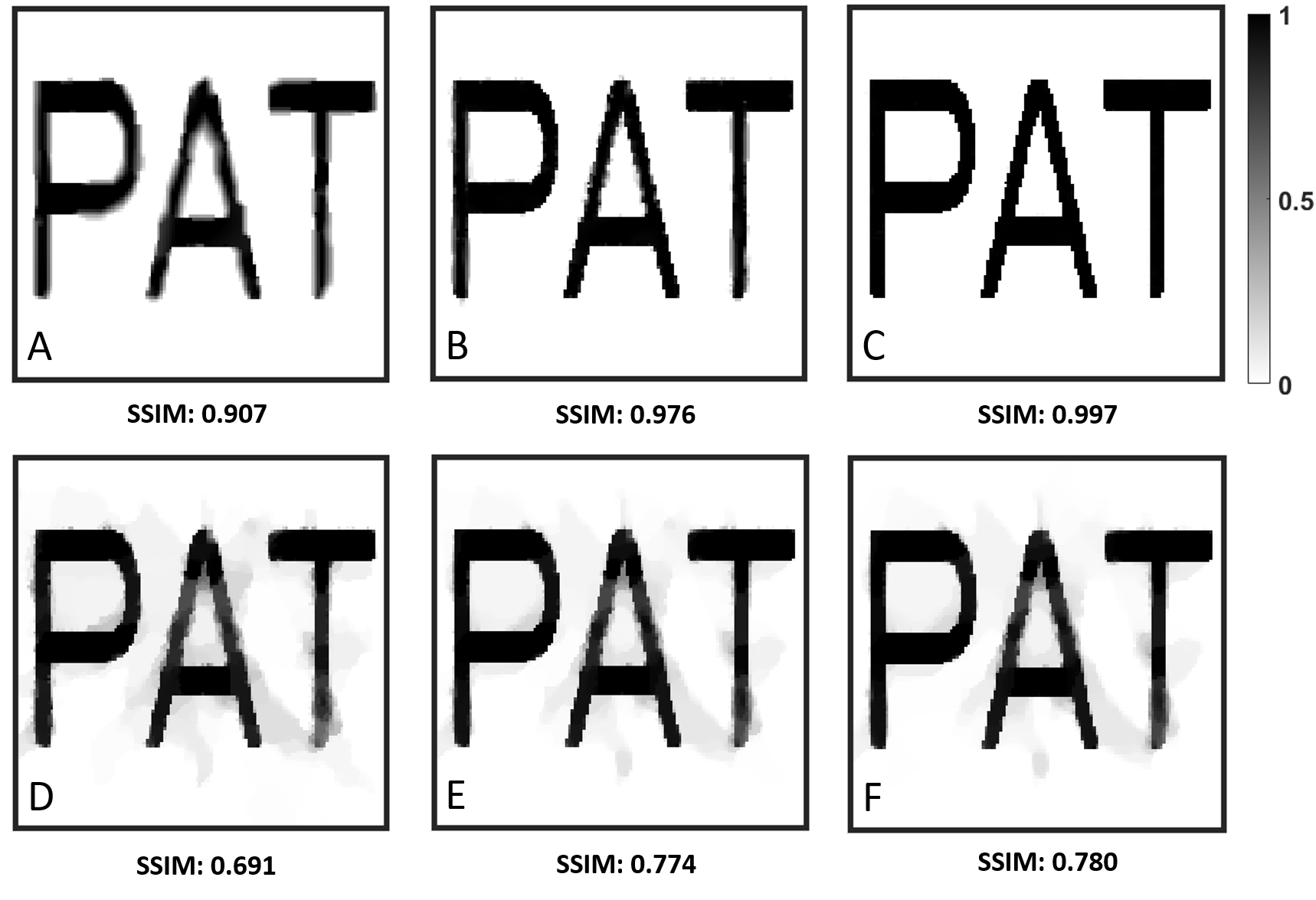}}
			\caption{Reconstructed images from input data sets corresponding to 16 transducers for various noise levels. (A),  (B), and  (C):   reconstructions obtained from the proposed method with second form of regularization corresponding to SNRs 20 dB, 30 dB, and 40 dB. (D),  (E), and  (F):   reconstructions obtained
				from the FISTA method corresponding to SNRs 20 dB, 30 dB, and 40 dB.}
			\label{fig:reccomp10}
		\end{figure}
		
		\begin{table*}[ht]
			\centering
			\begin{tabular}{ |c| c| c|c| c | c| c| c|c| c| c| c|c|}
				\hline
				\multicolumn{5}{|c|}{Blood Vessel Phantom} &\multicolumn{4}{|c|}{Derenzo Phantom} & \multicolumn{4}{|c|}{PAT Phantom}
				\\\cline{1-13}
				\multirow {2}{*} { SNR }  & \multicolumn{2}{|c|}{64 trans.} & \multicolumn{2}{|c|}{128 trans.} & \multicolumn{2}{|c|}{16 trans.} & \multicolumn{2}{|c|}{32 trans.} & \multicolumn{2}{|c|}{16 trans.} & \multicolumn{2}{|c|}{32 trans.} \\\cline{2-13}
				(dB) & FBM  & Ours  &FBM & Ours & FBM  & Ours  & FBM & Ours  & FBM  & Ours  &FBM & Ours\\
				\hline
				20 &  .640 & .758 & .705 & .762 &  . 682 & .983 & .914 & .994  &  .691 & .907 & .805 & .957\\
				\hline
				30 & .723 & .762 & .737 & .763 & .725 & .997 & .982 & .999  &  .774 & .976 & .976 & .999\\
				\hline
				40 & .743 & .762 & .742 & .764 & .740 & .999 & .989 & .999  &  .780 & .997 & .985 & .999\\
				\hline
			\end{tabular}
			\caption{ SSIM scores of reconstruction with varying input noise levels and different number of transducers for all phantoms.
				FBM:  FISTA-based method of Huang et al.  \cite{huang2013full} }
			\label{table:1}
		\end{table*}

		\subsection{Reconstruction from real measured data}
		
		We use a triangular-shaped physical phantom constructed using horsehair to generate real measured data.  The details of the experimental setup can be found in Fig.2 of Ref. \cite{awasthi2018image}. A Q-switched Nd: YAG   laser operating at 532 nm delivered laser pulses with 5 ns width at a 10 Hz repetition rate onto the sample. One uncoated plano-concave lens (LC1715, Thorlabs) and four right-angle uncoated prisms (PS911, Thorlabs) were utilized to provide a light fluence of 9  mJ/cm$^{2}$ ($<$ 20 mJ/cm$^{2}$: ANSI safety limit). The hair phantom having the side-length and diameter of 10 and 0.15 mm, respectively, was attached to the pipette tips adhered to an acrylic slab. For recording the PA data, a 2.25 MHz flat ultrasound transducer  (Olympus-NDT, V306-SU) of 13 mm diameter active area and  $  70 \% $ nominal bandwidth was rotated continuously for 360 deg around the sample. A pulse amplifier  (Olympus-NDT, 5072PR) first amplified and filtered the acquired PA signals, and then a data acquisition (DAQ) card (GaGe, compuscope 4227) recorded the signals using a  sampling frequency of 25 MHz. A sync signal from the laser was used for the synchronization of data acquisition with laser illumination.

		\begin{figure}[H]
			\centering
			\fbox{\includegraphics[width=0.65\linewidth]{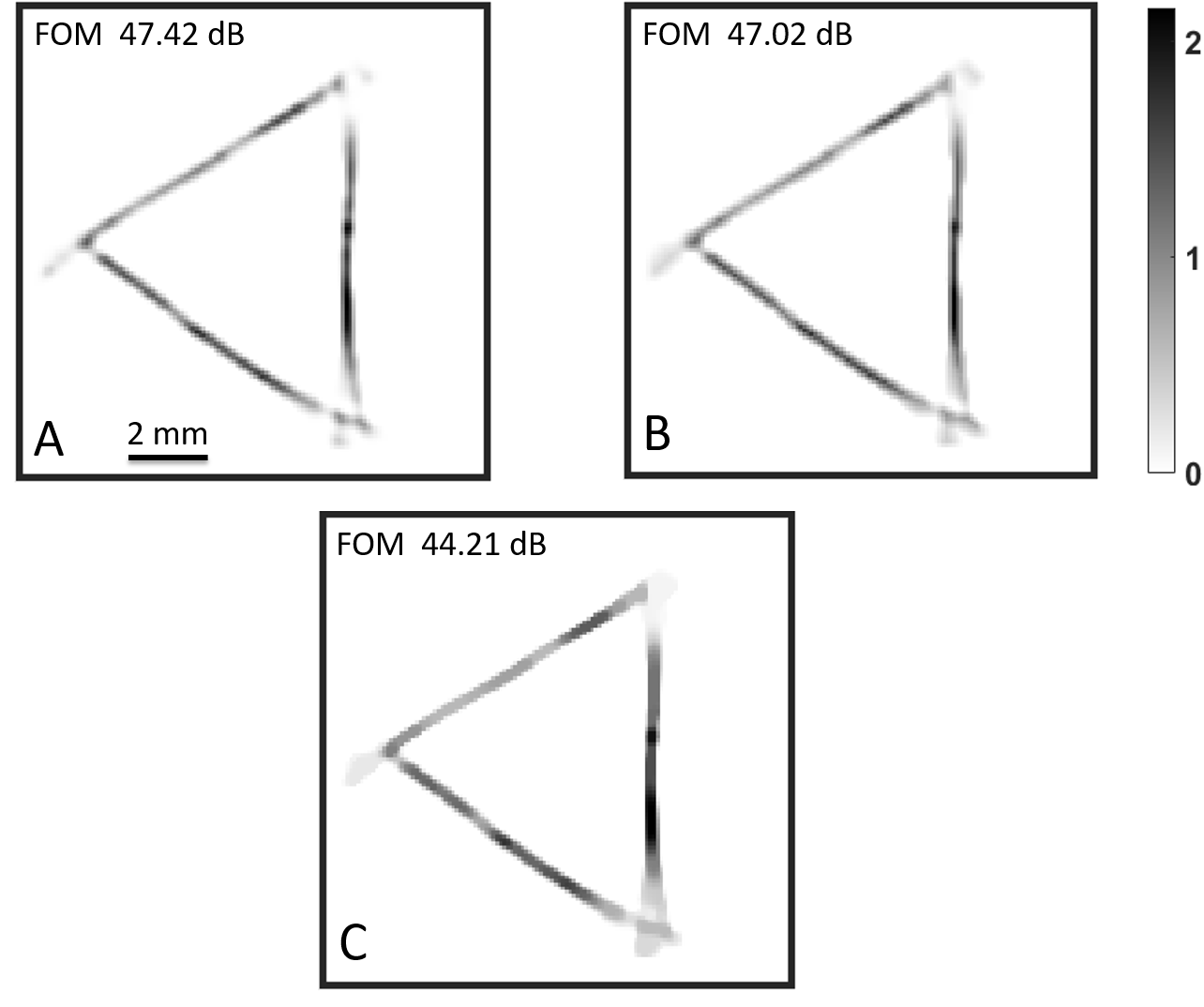}}
			\caption{Reconstructed images from horse hair phantom data using 400 transducers.
				(A):  reconstruction from the proposed method with the first form of regularization (FOM 47.42 dB).
				(B):  reconstruction from proposed method with second form of regularization 
				(FOM 47.02 dB); (C):  reconstruction from  FISTA-based method (FOM 44.21 dB).}
			\label{fig:recrealdata}
		\end{figure}

		The reconstructed PA imaging region has a size of  12.8 mm by 12.8 mm containing 128 by 128 pixels, and data from 400 transducer positions are used to do the reconstruction.
		Since the actual values of the initial pressure rise are unknown here, 
		we have used the following figure-of-merit as used in the reference \cite{li2017multiview}
		to compare different methods.
		
		\begin{equation}
			\label{eq:fomdef}
			FOM = 20 \times log_{10} \left( \cfrac{S}{n} \right)
		\end{equation}
		where $S$ is the peak intensity value of the reconstructed image and $n$ is the standard 
		deviation of the intensity. 
		
		Figure \ref{fig:recrealdata}  compares reconstructed images for  horse hair phantom data. For the proposed method, since we do not have the ground truth for evaluation, we present reconstruction results using both regularization forms. Figure  \ref{fig:recrealdata}.A    shows the image reconstructed using the first form of regularization term, and  Figure  \ref{fig:recrealdata}.B shows the result obtained by using the second form regularization.    Figure  \ref{fig:recrealdata}.C shows the corresponding result obtained from the FISTA based method. The proposed methods were able to give sharp images while giving a  3dB improvement in the FOM values compared with the FISTA based result, as given in the figure.

		\section{Conclusions}
		
		
		
		A novel model-based method that can reconstruct high-quality PAT images from significantly reduced measurements was developed. The joint sparse nature of intensity and derivatives in PAT images was exploited to construct the regularization. The reconstruction was formulated as the minimizer of this regularization along with data fidelity cost and a cost that quantifies the deviation from positivity. A novel computational method was developed to carry out the minimization.  Further, a new computational formula was derived for implementing the forward model and was used in the data fidelity term to reduce the memory requirement. The algorithm was compared against the FISTA based method of Huang et al. \cite{huang2013full} for various levels of reduction in the measured data  (16, 32, 64, and 128 transducers) with various levels of measurement noise (20, 30, and 40 dB).  We considered both real and simulated data sets for our experiments, and the proposed method yielded superior reconstruction quality in all cases.

		\bibliography{rejesh}

\providecommand{\newblock}{}
\begin{thebibliography}{10}
\expandafter\ifx\csname url\endcsname\relax
  \def\url#1{{\tt #1}}\fi
\expandafter\ifx\csname urlprefix\endcsname\relax\def\urlprefix{URL }\fi
\providecommand{\eprint}[2][]{\url{#2}}

\bibitem{wang2016practical}
Wang L~V and Yao J 2016 {\em Nature methods\/} {\bf 13} 627

\bibitem{pramanik2008design}
Pramanik M, Ku G, Li C and Wang L~V 2008 {\em Medical physics\/} {\bf 35}
  2218--2223

\bibitem{beard2011biomedical}
Beard P 2011 {\em Interface focus\/} {\bf 1} 602--631

\bibitem{zhou2016tutorial}
Zhou Y, Yao J and Wang L~V 2016 {\em Journal of biomedical optics\/} {\bf 21}
  061007

\bibitem{upputuri2016recent}
Upputuri P~K and Pramanik M 2016 {\em Journal of Biomedical Optics\/} {\bf 22}
  041006

\bibitem{li2017single}
Li L, Zhu L, Ma C, Lin L, Yao J, Wang L, Maslov K, Zhang R, Chen W, Shi J {\em
  et~al.\/} 2017 {\em Nature biomedical engineering\/} {\bf 1} 0071

\bibitem{Han:17}
Han Y, Ding L, Ben X~L~D, Razansky D, Prakash J and Ntziachristos V 2017 {\em
  Opt. Lett.\/} {\bf 42} 979--982

\bibitem{rosenthal2013acoustic}
Rosenthal A, Ntziachristos V and Razansky D 2013 {\em Current medical imaging
  reviews\/} {\bf 9} 318--336

\bibitem{finch2004determining}
Finch D and Patch S~K 2004 {\em SIAM journal on mathematical analysis\/} {\bf
  35} 1213--1240

\bibitem{kunyansky2007explicit}
Kunyansky L~A 2007 {\em Inverse problems\/} {\bf 23} 373

\bibitem{xu2002time}
Xu M and Wang L~V 2002 {\em IEEE transactions on medical imaging\/} {\bf 21}
  814--822

\bibitem{xu2005universal}
Xu M and Wang L~V 2005 {\em Physical Review E\/} {\bf 71} 016706

\bibitem{kalva2016experimental}
Kalva S~K and Pramanik M 2016 {\em Journal of Biomedical Optics\/} {\bf 21}
  086011

\bibitem{pramanik2014improving}
Pramanik M 2014 {\em JOSA A\/} {\bf 31} 621--627

\bibitem{xu2004time}
Xu Y and Wang L~V 2004 {\em Physical review letters\/} {\bf 92} 033902

\bibitem{burgholzer2007exact}
Burgholzer P, Matt G~J, Haltmeier M and Paltauf G 2007 {\em Physical Review
  E\/} {\bf 75} 046706

\bibitem{treeby2010photoacoustic}
Treeby B~E, Zhang E~Z and Cox B~T 2010 {\em Inverse Problems\/} {\bf 26} 115003

\bibitem{hristova2008reconstruction}
Hristova Y, Kuchment P and Nguyen L 2008 {\em Inverse Problems\/} {\bf 24}
  055006

\bibitem{xu2004reconstructions}
Xu Y, Wang L~V, Ambartsoumian G and Kuchment P 2004 {\em Medical physics\/}
  {\bf 31} 724--733

\bibitem{arridge2016accelerated}
Arridge S, Beard P, Betcke M, Cox B, Huynh N, Lucka F, Ogunlade O and Zhang E
  2016 {\em Physics in Medicine \& Biology\/} {\bf 61} 8908

\bibitem{buehler2011model}
Buehler A, Rosenthal A, Jetzfellner T, Dima A, Razansky D and Ntziachristos V
  2011 {\em Medical physics\/} {\bf 38} 1694--1704

\bibitem{paltauf2002iterative}
Paltauf G, Viator J, Prahl S and Jacques S 2002 {\em The Journal of the
  Acoustical Society of America\/} {\bf 112} 1536--1544

\bibitem{schwab2018galerkin}
Schwab J, Pereverzyev~Jr S and Haltmeier M 2018 {\em SIAM Journal on Numerical
  Analysis\/} {\bf 56} 160--184

\bibitem{jp2019}
{Prakash} J, {Mandal} S, {Razansky} D and {Ntziachristos} V 2019 {\em IEEE
  Transactions on Biomedical Engineering\/}  1--1 ISSN 0018-9294

\bibitem{huang2013full}
Huang C, Wang K, Nie L, Wang L~V and Anastasio M~A 2013 {\em IEEE Transactions
  on Medical Imaging\/} {\bf 32} 1097--1110

\bibitem{arridge2016adjoint}
Arridge S~R, Betcke M~M, Cox B~T, Lucka F and Treeby B~E 2016 {\em Inverse
  Problems\/} {\bf 32} 115012

\bibitem{boink2018framework}
Boink Y~E, Lagerwerf M~J, Steenbergen W, van Gils S~A, Manohar S and Brune C
  2018 {\em Physics in Medicine \& Biology\/} {\bf 63} 045018

\bibitem{saratoon2013gradient}
Saratoon T, Tarvainen T, Cox B and Arridge S 2013 {\em Inverse Problems\/} {\bf
  29} 075006

\bibitem{han2015sparsity}
Han Y, Tzoumas S, Nunes A, Ntziachristos V and Rosenthal A 2015 {\em Medical
  physics\/} {\bf 42} 5444--5452

\bibitem{wang2013accelerating}
Wang K, Huang C, Kao Y~J, Chou C~Y, Oraevsky A~A and Anastasio M~A 2013 {\em
  Medical physics\/} {\bf 40}

\bibitem{rosenthal2010fast}
Rosenthal A, Razansky D and Ntziachristos V 2010 {\em IEEE transactions on
  medical imaging\/} {\bf 29} 1275--1285

\bibitem{rosenthal2012efficient}
Rosenthal A, Jetzfellner T, Razansky D and Ntziachristos V 2012 {\em IEEE
  transactions on medical imaging\/} {\bf 31} 1346--1357

\bibitem{liu2016curve}
Liu H, Wang K, Peng D, Li H, Zhu Y, Zhang S, Liu M and Tian J 2016 {\em IEEE
  transactions on medical imaging\/} {\bf 35} 2546--2557

\bibitem{arigovindan2013high}
Arigovindan M, Fung J~C, Elnatan D, Mennella V, Chan Y~H~M, Pollard M, Branlund
  E, Sedat J~W and Agard D~A 2013 {\em Proceedings of the National Academy of
  Sciences\/}  201315675

\bibitem{cox2007k}
Cox B~T, Kara S, Arridge S~R and Beard P~C 2007 {\em The Journal of the
  Acoustical Society of America\/} {\bf 121} 3453--3464

\bibitem{cox2005fast}
Cox B and Beard P 2005 {\em The Journal of the Acoustical Society of America\/}
  {\bf 117} 3616--3627

\bibitem{xu2006photoacoustic}
Xu M and Wang L~V 2006 {\em Review of scientific instruments\/} {\bf 77} 041101

\bibitem{wang2011imaging}
Wang K, Ermilov S~A, Su R, Brecht H~P, Oraevsky A~A and Anastasio M~A 2011 {\em
  IEEE transactions on medical imaging\/} {\bf 30} 203--214

\bibitem{wang2012investigation}
Wang K, Su R, Oraevsky A~A and Anastasio M~A 2012 {\em Physics in Medicine \&
  Biology\/} {\bf 57} 5399

\bibitem{tao2009alternating}
Tao M, Yang J and He B 2009 {\em TR0918, Department of Mathematics, Nanjing
  University\/}

\bibitem{wang2008new}
Wang Y, Yang J, Yin W and Zhang Y 2008 {\em SIAM Journal on Imaging Sciences\/}
  {\bf 1} 248--272

\bibitem{nestedcg}
{Skariah} D~G and {Arigovindan} M 2017 {\em IEEE Transactions on Image
  Processing\/} {\bf 26} 4471--4482

\bibitem{nikolova2010}
{Nikolova} M, {Ng} M~K and {Tam} C 2010 {\em IEEE Transactions on Image
  Processing\/} {\bf 19} 3073--3088

\bibitem{ssim}
{Zhou Wang}, {Bovik} A~C, {Sheikh} H~R and {Simoncelli} E~P 2004 {\em IEEE
  Transactions on Image Processing\/} {\bf 13} 600--612

\bibitem{awasthi2018image}
Awasthi N, Kalva S~K, Pramanik M and Yalavarthy P~K 2018 {\em Journal of
  Biomedical Optics\/} {\bf 23} 091413

\bibitem{li2017multiview}
Li L, Zhu L, Shen Y and Wang L~V 2017 {\em Journal of biomedical optics\/} {\bf
  22} 076017

\end{thebibliography}
		\bibliographystyle{iopart-num}
		
	\end{document}